\documentstyle[12 pt]{article}
\begin{document}
\begin{center}
\null
\rightline {UFR-HEP/2000-1}
\vskip1truecm 
{\large\bf  Hyperkahler Singularities in Superstrings 
Compactification 
and $2d$ $N=4$ Conformal Field Theory}\\
\vskip1truecm  
 A. Belhaj $^1$ and E.H.Saidi$^{1,2}$
\end{center} 
\small $^1$  High Energy Physics Laboratory, Faculty of 
Sciences,  Rabat, Morocco.\\
 \small $^2$ International Institute of Theoretical and 
Applied Physics; IITAP,Ames.Iowa.USA\\
\begin{center} 
H-saidi@fsr.ac.ma;  b-adil@emi.ac.ma
\end{center}  
\vskip1truecm
\begin{abstract}
We study the singularities  of the Higgs branch of 
supersymmetric $U(1)^r$ gauge theories with eight supercharges. 
We derive new solutions for the moduli space of vacua 
preserving manifestly the eight supercharges by using a 
geometric realization of the $SU(2)_R$ symmetry and a 
separation procedure of the gauge and $SU(2)_R$ charges, 
which allow us to put the hypermultiplet vacua in a form 
depending on a parameter $\gamma$. For $\gamma=1$, we 
obtain new models which flow in the infrared to $2d$ 
$N=(4,4)$ conformal models and we show that the classical 
moduli spaces are given by intersecting cotangent weighted 
complex projective spaces containing the small instanton 
singularity, discussed in [17], as a leading special case. 
We also make comments regarding the $2d$ $N=4$ conformal 
Liouville description of the Higgs branch throat by following 
the analysis of [18]. Other features are also discussed.   
\end{abstract}	
\newpage
\section{Introduction}

\qquad Over the few past years, there has been an increasing 
interest in studying the moduli space of vacua of the Coulomb 
and Higgs branches of supersymmeric gauge theories with eight 
supercharges in various dimensions. This interest is mostly 
due to the fact that extended supersymmetry severely restricts 
the quantum corrections to the moduli space metric and allows 
to make many exact computations [1,2,3]. A large class of 
these gauge theories can be realized by brane configurations 
in type II strings on Calabi Yau manifolds  by using either 
the Hanany -Witten method [4,5] or  the geometric engineering 
approach introduced and developed by Klemm ,Lerche , Mayr, 
Vafa and Warner [6] and their collaborators; 
see   [7,8,9,10,11]; see also [12,13,14,15,16].\\ 
Recently, a special interest has been  given to the 
analysis of the hypermultiplet gauge invariant moduli 
space near the Higgs branch singularity. This analysis 
has been shown to be relevant for the study of many 
aspects in supersymmetric gauge theories with eight 
supercharges; in particular in the understanding of 
the assymptotic regions of the  infrared IR low energy 
limits of the  $N=(4,4)$ supersymmetric gauge theories in 
two dimensions especially within the so called throats of 
the Coulomb and  Higgs branches where the theories are 
typically described by two $2d$ $N=4$ conformal Liouville 
theories with different central charges. In this context, 
it was shown in [17]; see also [18,19,20,21,22,23], that 
the IR limits of the Coulomb and Higgs branches have 
isomorphic throat regions associated with different small 
$2d$ $N=4$ subalgebras of the  $N=4$ superconformal symmetry 
in two dimensions [17,18,19,20] see also [24-28].\\ Vector and 
hypermultiplet moduli spaces have been also much studied in 
strings compactification to four dimensions where the low 
energy supergravity has scalar fields in both vector multiplets
 and hypermultiplets. The basic example of such 
compactification is given by  type IIA string on 
$ R^{1,3}$ times Calabi Yau threefolds which is believed to 
be dual to heterotic on $R^{1,3}\times K3\times T^2$, where 
the type IIA dilaton is in a hypermultiplet and the heterotic 
one is in a vector multiplet [29,30]. Using local mirror 
symmetry and toric geometry methods, the absence of the 
type IIA dilaton in the vector multiplet has been exploited
 in [8]to derive exact results in the Coulomb branch of type 
IIA on local Calabi Yau threefolds. In the same spirit, the 
hypermultiplet moduli space is independant of the heterotic 
string coupling and hence can be determined exactly from 
heterotic conformal field theory near the $ K3 $ ADE 
singularity. This issue has been analysed recently in [31] 
and it was suggested that, in absence of small instanons, 
the hyperKahler moduli space for the heterotic string near the 
$ K3 $ ADE singularity is just the moduli space of vacua of a 
pure supersymmetric gauge theory in three dimensions with eight 
supercharges and ADE gauge group [32,33]. Matter adjunction has
 been considered in [34,35]. In the presence of small 
instantons, the hyperKahler moduli space has singularities 
which generally are interpreted in terms of singular conformal 
field theories or non perturbative massless particles. Note 
that contrary to the singularity of the coulomb branch 
generated by the one loop quantum corrections, the singularitiy 
of the Higgs branch of supersymmetric gauge theories with eight 
supercharges is not generated by quantum mechanics. It has 
been first  suspected when tempting to understand the 
breakdown of string perturbation theory  in type IIA on $A_r$ 
ALE surface, where the non perturbative phenomenon cannot be 
avoided by making the string coupling constant smaller 
[17,36,37]. In lower dimensions, the Higgs branch singularity 
was also motivated  by using duality between Higgs and Coulomb 
of $ N=(4,4)$ supersymmetric gauge theories in two dimensions 
[22,23]. More convincing  and rigourous arguments for the 
existence of the Higgs branch throat are obtained from the 
study of the low energy physics of the D1/D5 system on $X$, 
which is equivalent to type IIB String theory on 
$AdS_3\times S^3\times X$ where $X$ is either $T^4$ or 
$K3$ [17,18,20,38].\\
 In four dimensions, hypermultiplets moduli are moreover 
involved in the study of stringy instanton moduli space 
which is given by hyperkahler deformations  of the classical 
instanton moduli space with non zero B field and small 
instanton singularites eliminated. Stringy instantons with 
non zero B field, including hyperkahler deformations  
resolving small instanton singularities, have been suggested 
in [39] to be equally described as instantons on a non 
commutative space [40,41 ]. Furthermore analysis involving 
hypermultiplets moduli is also encountered in the study of  
CFT's obtained  from  compactifications of superstrings on 
Calabi Yau fourfolds and too particularly in the 
compactification of M-Theory on a Calabi Yau four-folds 
near the so called hyperkahler singularity [42].\\
 In most of  all of these studies, one  of the basic eqs 
describing the moduli space of the hypermultiplets vacua reads,
 in the sigma model approach, as:
\begin{equation}
\label{1}
\sum_{i=1}^{n} q_a^i[\varphi_i^\alpha \overline{\varphi}_{i\beta}+
\varphi_{i\beta}\overline{\varphi}_i^\alpha]=
\vec{\xi_a} \vec{\sigma}^\alpha_\beta\qquad;a=1,...,r.
\end{equation}
Other basic eqs describing  the throat region of the Higgs branch  
are given in section 5; see eqs(65,66). 
In  eqs(1), the $\varphi^\alpha_i$'s form a set 
$\{\varphi^\alpha_i;1\leq i\leq n\}$  of  $n$ component 
fields doublets $\varphi^\alpha_i$ belonging to  hypermultiplets 
and transforming in the $(n,2)$ representations of $G\times SU(2)_R$ 
group, where  the group G will be specified later on. The $\vec{\xi}_a$'s 
are a collection of $r$ FI coupling 3-vectors, each of it transforms as a 
triplet under the usual $SU(2)_R$  symmetry rotating the eight supercharges. 
The $q^i_a$ parameters are the charges of $\varphi^\alpha_i$'s under the $ U(1)^r$  
gauge group of the underlying supersymmetric gauge theory. For later use it is 
interesting to note that eqs(1) have a formal analogy with the following sigma model 
vaccum eqs of $2d$ $ N=2 $  supersymmetric $U(1)^r$ gauge  theory involved in the 
analysis of the Coulomb branch of IIA superstrings on Calabi Yau threefolds with ADE 
singularities
\begin{equation}
\label{2}
\sum_i q_a^i|X_i|^2=R_a; \qquad a=1,...,r. 
\end{equation}
In eqs (2), the $X_i$'s are complex scalar fields , the $R_a$'s are FI 
couplings and  the $q_a^i$'s are the $U(1)^r$ charges of the $X_i$'s which, 
for reference , read in the case of $SU(n)$ singularity as :
\begin{equation}
\label{3}
\ q_a^i=-2\delta_a^i+\delta_a^{i-1}+\delta_a^{i+1},
\end{equation}
with the remarkable equality
\begin{equation}
\label{4}
\sum_i q_a^i=0.
\end{equation} 
It is interesting to note here that in the $2d$ gauge theories with four 
supercharges, the above constraint eqs (4) is the condition under which 
the gauge theory flows in the infrared to $2d$ $ N=2$ superconformal field 
theory [43,44]. It is also the condition to have Kahler  Calabi Yau backgrounds 
[8,45] involved in superstring compactifications. Concerning eqs(1), we will see 
in section 4 that, under some assymptions, one can works out two remarkable  
classes of gauge invariant solutions of eqs (1) preserving the eight supercharges. 
The first class leads to the obtention of new singularities  extending the usual $N=2$ 
ADE ones which are recovered as a special solutions by partial breaking of $2d$ $N=4$ 
supersymmetry down to $2d$ $N=2$. As preliminary results we find the following singular
 hypersurfaces:
\begin{equation}
\label{5}
\begin{array}{lcr}
A_{n-1}:\qquad U^{+{n(n+1)\over2}}V^{+{n(n+1)\over2}}=[Z^{+{(n+1)}}]^n\\
D_{n}:\qquad(x^{++})^{n}+x^{++}(y^{+(n-1)})^2+(z^{+n})^{2}=0\\
 E_6:\qquad (x^{+6})^2+(y^{+4})^3+(z^{+3})^4=0\\
 E_7:\qquad (x^{+9})^2+(y^{+6})^3+y^{+6}(z^{+4})^3=0\\
 E_8:\qquad (x^{+15})^2+(y^{+10})^3+(z^{+6})^5=0.
\end{array}
\end{equation}
In these eqs the charges carried by the various gauge invariants $U,V,Z,x,y$ and $z$  
are Cartan charges of the $U(1)_R$ abelian subsymmetry of $SU(2)_R$ group. Eqs (5) 
extend the usual ADE complex surfaces of the $N=2$  backgrounds. For more details, 
see eqs (35-36). In the second class of gauge invariant solutions, we will show that  
eqs (4) are no longer constraint eqs; they are replaced by the following remarkable 
identity:
\begin{equation}
\label{4}
\sum_i q_a^i +\sum_i (-q_a^i)=0, 
\end{equation}
which is usually  fulfilled whatever the values of the $q^i_a$'s are.
In the low energy limit,  the above eqs lead then to $2d$ $ N=4$ conformal 
models going beyond the  $ N=2$ ADE  conformal ones.\\  
The resemblance between eqs (1) and (2) is  only formal, but turns out however  
to be very useful for the analysis of eqs(1) as well as their solving. Eqs(1) and (2) 
carry different meanings amongst which  we quote the four followings:\\ (a) Eqs (2) 
describe the Coulomb branch leading to the  well known   gauge invariant  Kahler 
moduli space while eqs (1) deal with the hypermultiplet branch and give a gauge 
invariant hyperkahler  moduli space.\\ (b) For each $U(1)$ factor of the $U(1)^r$ 
gauge group, eqs (1) involve a triplet of FI parameters whereas eq (2) has only one. 
This feature goes with the previous one as it  is related with the number of complex 
stucutres of Kahler and hyperkahler manifolds. \\ (c) Eqs (1) have a manifest  $SU(2)_R$
 symmetry which is absent in eqs (2).  The latters have a $U(1)$ R symmetry. Eqs (1) 
are more restrictive than eqs(2) since they are the vaccum eqs of $2d$ $ N=4$ 
supersymmetric linear $\sigma$ models. More precisely $N=4$ models in two 
dimensions involve three times the number of the D-flatness eqs of the $N=2$ 
models. This feature is easily seen on the space of the  FI couplings which, 
for $N=4$, belong to $({R^3})^r\approx R_+^r\times ({S^2})^r$ while, for $N=2$, 
belong to $R^r$. The extra eqs associated with the moduli space $({S^2})^r$ are 
necessary conditions to have $N=4$ supersymmetry; without these constraints $ N=4$ 
supersymmetry is partially broken down to $N=2$.\\ (d) It is now quite well 
established that eqs (1) and (2) hide moreover a comparable behaviour between 
the Coulomb and Higgs branches even near the singularity where eqs (1) and eqs (2) 
ceasse to be valid. We have already mentioned the duality  between the two branches 
and their algebraic descriptions in terms of subalgebras of $2d$ $N=4$ conformal 
invariance. Later on, we shall give other arguments, geometrical and field theoretical, 
showing that in absence of FI couplings, $\theta$ terms and  RR fields , both Coulomb 
and Higgs branches are described by singular CFT's which seems to have something to do 
with $2d$ $ N=4$ superconformal ADE Toda theories.\\   
The formal similarity between eqs (1) and (2) together with the abovementioned 
features show that one may obtain  new solutions of the gauge invariant moduli 
space of vacua of eqs (1) by using $SU(2)_R$ harmonic analysis and generalisations 
of methods of $2d $ $N=2$ supersymmetric linear sigma models. \\
The aim of this paper is to study these solutions and give interpretations in terms 
of blown up of  singularities given by intersections of cotangent complex $n$ 
dimensional weighted projective spaces. Actually this  study extends the results 
obtained  for Coulomb branch of supersymetric gauge theories  with four supercharges. 
Concerning  the infrared dynamics of two dimensional $N = (4,4)$ gauge theories, we 
give also comments on the $ N=4$  conformal Liouville description of the region in 
the viccinity of the singularity of the metric of the  $2d $ $N=4$ Higgs branch 
generally interpreted as a semi infinite throat where the string coupling constant 
$g_s=e^\phi$ blows up as the Liouville field $\phi$ goes to infinity [17,18].  
Moreover, in an attempt towards an interpretation of the degenerate  $ A_r$ 
singularity  carried by eqs (1), we give  a field theoretical argument in favor to 
the hypothesis according to which the metric of the moduli space near the Higgs 
singularity  maight be described  by a $N=4$ conformal $SU(r+1)$ Toda theory in 
two dimensions. Of course this is just an observation which deserves  in its own 
right a detailed  study.
 \\ The presentation of this paper is as follows: In section 2, we review brefly 
the standard way  used in handling eqs (1) where only a $A_1$ singularity has been 
considered, using the standard $2d$ $ N=2$ supersymmetric analysis in which  half 
of the eight supersymmetries are manifest. In this way of doing the $SU(2)_R$ 
symmetry is broken down to $U(1)_R$, a feature which is exploited  in  [31,42] 
by making an appropriate choice  of the FI 3-vector coupling where only one 
parameter is non zero. The two others are put to zero. In section 3, we develop a  
new way  of doing  by keeping all the three FI parameters non zero and the eight 
supercharges  manifest. In this approach  the $SU(2)_R$ symmetry is apparent but 
explicitly broken by the non zero FI  terms. Our method enables us to exhibit 
manifestly the  role of the three Kahler  structures of the gauge invariant 
hyperkahler moduli spaces and permits moreovver to go beyond the $A_1$ singularity 
analysis of [31,42 ]. Our way in handling eqs(1) involves two steps  based  on a 
geometric realisation of the $SU(2)_R$  symmetry and on the separation of the 
charges of the gauge and $R$-symmetries. The first step of this programme is 
described in Sectin 3 while the second step is studied in Section 4. The gauge 
and $SU(2)_R$ charge separation of the hypermultiplets moduli involves  a parameter 
$\gamma$  taking the values $\gamma=0$ or $\gamma=1$ which distinguish two classes 
of solutions of eqs(1) both preserving the eight supercharges. For $\gamma=0$, we 
obtain a generalisation of the ADE complex surfaces reproducing the standard ones 
by partial breaking of $2d$ $N=4$ supersymmetry down to $2d$ $N=2$. For $\gamma=1$, 
we find new models which flow in the infrared to $2d$ $N=(4,4)$ scale invaraint models. 
In section 5, we study the moduli space of vacua of models with $\gamma=1$ by 
distinguishing the two cases $\sum\limits _iq^i_a=0$ and  $\sum\limits _iq^i_a \ne 0$.
 We show by explicit computation that the hyperKahler moduli space, associated with 
eqs (1), is given  by weighted complex projective spaces. Moreover, we study the 
Liouville description of the small instanton conformal theory near the singularity 
by using the field theoretical approach of Aharony and Berkooz [18] and make comments 
regarding the $A_r$ singularity of eqs (1). In  section 6, we give our conclusion.  
\section{Hyperkahler moduli space}
\qquad In this section we review brefly  the example of the hyperkahler cotangent
 bundle of complex projective space: $T^*(CP^2)$; considered in the study of M 
theory on Calabi Yau fourfold after what we give the 2n complex dimension hyperkahler 
space $T^*(CP^n), n\geq1$, describing  the instantons moduli space  of one instanton 
on $R^4$ with gauge group $U(n)$ [39,46]. To begin, note first of all that a Calabi 
Yau fourfolds can develop singularities  of many types; this includes the 
$C^4\over{Z_4}$ orbifold, the ADE  hypersurface singularities considered recently 
in [42] in the context of derivations of $2d  $ CFT's from type IIA string 
compactifications on Calabi Yau fourfolds , and the so called hyperkahler 
singularity we are intersted in here . To describe the $T^*(CP^2)$ bundle, 
we consider $2d $ $ N =4$ supersymmetric $U(1)$ gauge theory with one isotriplet 
FI coupling $\vec{\xi}=(\xi^1,\xi^2,\xi^3)$ and three hypermultiplets of charges 
$q^i_a=q^i=1;  i=1,2,3$ and taking G as $SU(3)$. The zero energy states of this 
gauge model are obtained by solving 
\begin{equation}
\label{6}
\sum_{i=1}^{3}[ \overline{\varphi}_{i\alpha} {\varphi}_i^\beta+
\varphi_{i\alpha}\overline{\varphi}_i^\beta]=\vec{\xi}\vec{\sigma}_\alpha^\beta,
\end{equation} 
which by the way is just  a special situation of eqs (1) where all gauge charges 
$q^j_a$ are equal to one. Eqs (6) is a system of three eqs which,up to replacing 
the Pauli matrices by their expressions and using the $SU(2)_R$ transformations 
$\varphi^{\alpha}=\varepsilon^{\alpha\beta}\varphi_ {\beta}$ with 
$\varepsilon_{12}=\varepsilon^{21}=1$ and 
$\overline{(\varphi^\alpha)}=\overline{\varphi}_\alpha$, split as follows:

\begin{equation}
\label{5}
\begin{array}{lcr}
\sum\limits_{j=1}^3( |\varphi^1_j|^2-|\varphi^2_j|^2) &= \xi^3 \qquad &
(a)\\ 
\sum\limits_{j=1}^3 \varphi^1_j \overline{\varphi}_{j2}&=\xi^1+i{\xi^2}\qquad 
&(b)
\\ 
\sum\limits_{j=1}^3 \varphi^2_j \overline{\varphi}_{j1}&=\xi^1-i{\xi^2}\qquad 
&(c).
\end{array}
\end{equation}
The moduli space of zero energy states of the classical gauge theory is the space 
of the solutions of eqs (7-8) divided by the action of the $U(1)$ gauge group. The 
solutions of eqs (7) depend on the values of the FI couplings. For the case where 
$\xi^1=\xi^2=\xi^3=0$, the moduli space has an $SU(3)\times SU(2)_R$ symmetry; it 
is a cone over a seven manifold described by the eqs:    
\begin{equation}
\label{7}
\sum_{i=1}^3(\varphi_{\alpha i}\overline{\varphi}_i^\beta-
\varphi_i^\beta\overline{\varphi}_{\alpha i})= \delta{ _\alpha ^\beta}.
\end{equation}
 For the case $\vec{\xi}\neq\vec{0}$, the abovementioned 
$SU(3)\times SU(2)_R$ symmetry is explicitly broken down to 
$SU(3)\times U(1)_R$. In the remarkable case where  $\xi^1=\xi^2=0$  and 
$\xi^3$ positive definite, it is not difficult to see that eqs (7) 
describe the cotangent bundle of $CP^2$. Indeed making the change
\begin{equation}
\label{9}
{\psi_i}={\varphi_{i}^{1}\over{[\sum\limits_{j=1}^3|\overline{\varphi}_{j2}|^2+
\xi^3]^{1\over2}}},
\end{equation}
and putting back into eq (8.a), one discovers that $\psi_{i}$'s 
satisfy $\sum_{i}{|\psi_i|^2}=1$. The $\psi_i$'s parametrize the 
$CP^2$ space. On the other hand with $\xi^1=\xi^2=0$  
 conditions, eqs (8.b-c) may be interpreted to mean that 
$ \overline{\varphi}_{2j}$ lies in the cotangent space to $CP^2$
 at the point determined by $\psi_i$.  Although we are usually 
allowed to make the choice $\xi^1= \xi^2=0$ by using an appropriate 
$ SU(2)_R$ transformation, we shall consider in sections 4 and 5,  
the generic  cases where $\xi^1, \xi^2$ and $\xi^3$ are all of them non 
zero as they form altogether the three Kahler parameters  of hyperkahler 
manifolds. For the time being let us note that the previous analysis may 
be extended to the cases of $2d$ $ N=4$ supersymmetric $U(1)$ gauge linear 
sigma model involving $n+1$ hypermutiplets with charges $q^i=1  ;  i=1,..., n+1 $ 
and transforming in the fundamental representation of $SU(n+1)$. The vaccum energy 
equations of this $U(1)$ gauge model read as :
\begin{equation}
\label{9}
\begin{array}{lcr}
\sum\limits_{j=1}^{n+1}( |\varphi_{j}^{1}|^2-|\overline{\varphi}_{2j}|^2)&=\xi^3\\
\sum\limits_{j=1}^{n+1}( \varphi_{j}^{1}\overline{\varphi}_{2j})&=\xi^1+i\xi^2.
\end{array}
\end{equation}
For $\xi^1= \xi^2=0$ and $\xi^3 $ positive definite, the  classical moduli 
space of the classical gauge theory is given by the cotangent bundle of 
complex $n$ projective space: $T^*CP^{n}$ . For $\xi^1= \xi^2=\xi^3=0$, 
one has just the conifold  singularity of $n$ dimensional complex manifolds. 
Note also that near this singularity the low energy limits of this gauge theory 
is described by 2d N=(4,4) superconformal field theory of central charge 
$ C=6(n+1-1)=6n$. In section 5, we shall turn to this point and describe the 
nature of the conformal field theory one has in the nearby of the Higgs branch 
singularity. 
\section{More on the Eqs (1)}
\qquad Eqs (1) is a system of $3r$ equations or more precisely $r$ isovector eqs, 
each of which shares some general features with the usual $2d$ $  N=2$ supersymmetic 
D-flatness conditions eqs (2); but in addition to the gauge charges, it carries a 
$SU(2)_R$ charge. To solve eqs (1), we shall use a different method than that used 
in [31,39,42]. This method   reproduces  the solutions of the abovementioned study 
as particular cases and offers moreover a possiblity to address the question of 
multi instantons in $ R^4$. Our approach is done in two steps and is based on the 
two following: First we combine methods of $2d $ $ N=2$  supersymmetric sigma models, 
as used in describing  the Kahler Coulomb branch of type IIA string on $K3$, 
and $SU(2)_R$ harmonic analysis  allowing us to interpret  $SU(2)_R$ 
representations as special functions on $ S_R^3=SU(2)_R$. The index R 
carried by $S_R^3$, $S_R^2$ and $U(1)_R$  refers to the $ SU(2)_R$. This 
step allows us to put eqs(1) into a manageable form which exhibit many 
similarities with eqs (2). Second , we introduce a convenient change of 
variables based on separating the $U(1)^r_G$ gauge charges and the  $U(1)_R$ ones. 
This change of variables allows us to benifit from the  similarities with the  
$2d$ $  N=2$ supersymmetric gauge invariant backgrounds in order  to study and 
solve eqs (1). In this section, we describe the first step of this programme; 
the charge separation of $U(1)_G$ and  $U(1)_R$  symmetry factors  will be 
studied in the next section. Our main purpose in what follows is to establish 
first that, up to $SU(2)_R$ transformations; eqs (1) can be rewitten in the 
following remarkable form:
\begin{equation}
\label{10}
\sum_{j}q^j_a \varphi_{j}^{+}\overline{\varphi}_{j}^{+}=-i\xi^{++}_a ;\qquad a=1,...,r;
\end{equation}
which , abstraction done of the plus indices describing the $U(1)_R$  Cartan 
charges carried by  the $\varphi _j$'s and the FI couplings, is comparable to eqs (2). 
In eqs (12); the moduli  $\varphi _j^+$, $\overline{\varphi } _j^+$ and $\xi_a^{++}$  
are related to $\varphi _j^\alpha$, 
$\overline{\varphi }_j^\alpha$ and $\xi_a^{(\alpha\beta)}$; 
they  will be specified later  on. To establish eqs (12)  
let first note that one may use the isomorphisms $ SU(2)_R=S_R^3$ and 
${SU(2)_R \over {U(1)_R}}\approx S_R^2$ to describe $SU(2)_R$ representations
 (both reducible and irreducible ) as harmonic functions on  the sphere $S_R^2$ 
 with definite $U(1)_R$ charge. This way of doing is well known in  the study of
 $SU(2)_R$ representation theory; the main idea behind this construction may be 
summarized as follows: First, consider the following  $2\times 2$ matrix       
\begin{equation}
\label{11}
U=\left(\matrix{
u^+_1&u^+_2\cr
u^-_1&u^-_2
\cr}\right)
\end{equation}
and solve the isospin $1\over{2}$ $SU(2)_R$ representation constraints 
namely the unimodularity $det U= 1$
and the unitarity $U^+U=U^+U=I$  conditions. Straightforward algebra 
leads to:[47,48,49]

\begin{equation}
\label{12}
\begin{array}{lcr}
u^{{\pm}{\alpha}}=\epsilon^{{\alpha}{\beta}}u^{\pm}_{\beta}; \qquad 
\overline{u}^{+\alpha}=u^-_{\alpha}; \qquad \epsilon_{\alpha\beta}=-
\epsilon_{\beta\alpha}\\

u^{+\alpha}u^{-}_{\alpha}=1,\qquad u^{+\alpha}u^{+}_\alpha =u^{-\alpha}u^{-}_\alpha=0.
\end{array}
\end{equation}
 
Recall in passing that the $u_\alpha^\pm$ harmonic variables are bosonic 
$SU(2)_R$ doublets which parametrize the unit  $S_R^3$ sphere; they may be 
solved in terms of the standard $S_R^3$ variables $\psi$, $\theta$ and $\phi$ as         
 \begin{equation}
\label{13}
\begin{array}{lcr}
u^+_1=\cos{\theta\over2} \exp{i\over2}(\psi+\phi)\\
u^+_2=\sin{\theta\over2} \exp{i\over2}(\psi-\phi)\\
u^-_1=\sin{\theta\over2} \exp{-i\over2}(\psi+\phi)\\
u^-_2=\cos{\theta\over2} \exp{-i\over2}(\psi-\phi).
\end{array}
\end{equation}
We shall not use this realization hereafter as we shall take   $u_\alpha^\pm$ 
as our basic variables. Moreover, using  the $u_\alpha^\pm$ variables,  
the $SU(2)_R$ algebra is realized as differential operators on the space 
of harmonic functions on $S_R^3$:
\begin{equation}
\label{14}
\begin{array}{lcr}
D^{++}=u^{+ \alpha}{\partial\over{\partial u^{-\alpha}}};\qquad D^{--}=
u^{- \alpha}{\partial\over{\partial u^{+\alpha}}}\\
2D^{++}=[D^0,D^{++}];\qquad -2D^{--}=[D^0,D^{--}]\\
D^0=[D^{++},D^{--}]=u^{+ \alpha}{\partial\over{\partial u^{+\alpha}}}-
u^{- \alpha}{\partial\over{\partial u^{-\alpha}}}
\end{array}
\end{equation}
To study the $SU(2)_R$ representations by using the harmonic variables, 
it is more convenient to consider harmonic functions $ F^q(u_\alpha^\pm)$  
with definte $U(1)_R$ charge q ; that is functions $ F^q(u_\alpha^\pm)$  
satisfying the eigenfunction eq 
\begin{equation}
\label{15}
[D^0,F^q]=qF^q.
\end{equation}
These functions  $F^q$ have a global harmonic expansion of total charge $q$ 
and carry $SU(2)_R$ representations. For example, taking $q=2$ and choosing  
$F^{++}$ as: 
\begin{equation}
\label{16}
F^{++}(u^{\pm}_\alpha)=u_{(\alpha}^+u_{\beta)}^+F^{(\alpha\beta)},
\end{equation}
one sees that $F^{++}$ is the highest state of the isovector representation of 
$SU(2)_R$. This is also seen from the following eqs defining the highest  
states of $SU(2)_R$ of $U(1)_R$ charge equal to $q$
$$[D^0,F^q]=qF^q$$
 \begin{equation}
\label{17}
\end{equation}
$$[D^{++},F^q]=0.$$
Thus the harmonic functions $F^{++}$  altogether with $F^0$  and $F^{--}$, 
defined as
$$F^0=[D^{--},F^{++}]=u_{(\alpha}^+u_{\beta)}^-F^{(\alpha\beta)}$$
 \begin{equation}
\label{18}
\end{equation}
$$F^{--}=[D^{--},F^0]=u_{(\alpha}^-u_{\beta)}^-F^{(\alpha\beta)},$$
form the three states of the isotriplet representation of the algebra (16) . 
In connection with the isotriplet representation $\{F^{++}$,$F^0$,$F^{--}\}$, 
there is an interesting feature that we want to give at this level and which we 
will use later on when studying the solutions of eqs (12). This feature concerns 
the fact that one can usually realize $F^q;  q=0, \pm2 $ as bilinears of 
isospinors $f^+$ and $\overline {f}^+$ as  follows
\begin{equation}
\label{19}
\begin{array}{lcr}
F^{++}=if^+\overline{f}^+=iu_{(\alpha}^+u_{\beta)}^+f^{(\alpha}\overline{f}^{\beta)}\\
F^0={i\over{2}}(f^+\overline{f}^-+f^-\overline{f}^+)={i\over2}u_{(\alpha}^+u_{\beta)}^-f^{(\alpha}\overline{f}^{\beta)}\\
F^{--}=if^-\overline{f}^-=iu_{(\alpha}^-u_{\beta)}^-f^{(\alpha}\overline{f}^{\beta)}.
\end{array}
\end{equation} 
The complex number  $i$ in front of the the factor of the right hand 
side of the above eqs ensures the reality condition of the isotriplet 
representation. Moreover the  realization of $F^q;  q=0, \pm2 $ as bilinears  
of isospinors reflects too simply the fact that the $ SU(2)_R$ isovectors may 
be built from  the symmetric product of the isospin $1\over2$  representation 
and its conjugate. After this digression on the $SU(2)_R$ harmonic analysis, 
we turn now to eqs (1) which we write as:
\begin{equation}
\label{20}
\begin{array}{lcr}
\sum\limits_{j}q^j_a \varphi_{j}^{+}\overline{\varphi}_{j}^{+}&=
-i\xi^{++}_a\qquad &(a) \\
\sum\limits_{j}q^j_a( \varphi_{j}^{+}\overline{\varphi}_{j}^-
+\varphi_{j}^{-}\overline{\varphi}_{j}^+)&=-2i{\xi^0_a}\qquad &(b)\\
\sum\limits_{j}q^j_a\varphi_j^-\overline{\varphi}_j^-&=
-i{\xi^{--}_a}\qquad &(c).
\end{array}
\end{equation} 
These eqs are obtained from eqs (1) by multipllying their both sides by 
$u_{(\alpha}^+u_{\beta)}^+$, $u_{(\alpha}^+u_{\beta)}^-$ and 
$u_{(\alpha}^-u_{\beta)}^-$ respectively. Eqs (22)  are also 
the D-flatness eqs one gets if one is using the $2d$ $ N=(4,4)$  
harmonic superspace formulation of $2d$ $N=4$ gauge theories [50,51]. 
Thus like for eqs (1), eqs (22) form altogether a system of $r$ isovector 
eqs of the $SU(2)_R$ algebra (16); but with the remarkable difference that 
now it is enough  to focuss attention on the highest weight states eqs (22.a). 
Knowing the solutions $\varphi_j^+$ and $ \overline{\varphi}_j^+$ of eqs (20-a), 
one can also get the solutions of $\varphi_j^-$ and $ \overline{\varphi}_j^-$ by 
acting on $\varphi^+$ and $ \overline{\varphi}^+$  by $ D^{--}$; namely:
\begin{equation}
\label{equation}
\begin{array}{lcr}
\varphi^-_j=[D^{--},\varphi^+_j] \\
\overline{\varphi}^-_j=[D^{--},\overline{\varphi}^+_j].
\end{array}
\end{equation}
In the end of this section, we would like to make two comments. 
The first comment is that one can use the isospinor bilinear realization 
of isotriplets eqs (21) to represent the Kahler parameters $\xi_a^{++}$ as follows:

\begin{equation}
\label{22}
\begin{array}{lcr}
\xi_a^{++}=i\zeta_a^+\overline{\zeta}^+_a=iu_{(\alpha}^+u_{\beta)}^+
\zeta^{(\alpha}_a{\overline{\zeta}^{\beta)}_a}\\
\zeta_a^{\pm}=u_{\alpha}^{\pm}\zeta^{\alpha}_a;\qquad\overline\zeta^{\pm}_a=
u_{\alpha}^{\pm} \overline{\zeta}^{\alpha}_a,
\end{array}
\end{equation}  
where $\zeta_a^\alpha$ and  $ \overline\zeta_a^\alpha$ may, roughly speaking, 
be viewed as  the square roots of the FI couplings $\xi_a^{(\alpha\beta)}$. 
Similar  relations involving $\zeta_a^{\pm}$ and  $ \overline{\zeta}_a^{\pm}$ 
for $\xi_a^0$ and  $ \xi_a^{--}$ may be also written down. Putting back these 
relations in eqs (22.a), one gets
\begin{equation}
\label{23}
\sum_{j}q^j_a \varphi_{j}^{+}\overline{\varphi}_{j}^{+}=\zeta_a^+\overline{\zeta}^+_a
=u_{(\alpha}^+u_{\beta)}^+\zeta^{(\alpha}_a\overline{\zeta}^{\beta)}_a.
\end{equation}
The second comment we want to do is that one can simplify further eqs (25) by making 
an extra change of variables which turns out to convenient when discussing the moduli 
space of gauge invariant vacua of eqs (1). This extra change consists to use the 
mapping $ R^3=R^+ \times S^2 $ to write the FI isovectors $\xi_a^{++}$ as
\begin{equation}
\label{24}
\xi_a^{++}= R_a\eta^+_a \overline{\eta}^+_a=r_a^2\eta^+_a \overline{\eta}^+_a,
\end{equation}
or equivalently by using the isospinors $\zeta_a^\alpha $ and  
$ \overline{\zeta}_a^{\alpha}$ introduced previously: 
\begin{equation}
\label{25}
\begin{array}{lcr}
\zeta^{\pm}_a&=u^{\pm}_\alpha \zeta^\alpha_a; \qquad \overline{\zeta}^{\pm}_a
=u^{\pm}_\alpha \overline{\zeta}^\alpha_a\qquad &(a)\\
\zeta^\alpha_a&=r_a\eta_a^\alpha;\qquad \overline{\zeta}^\alpha_a
=r _a\overline{\eta}_a^{\alpha}\qquad &(b)\\
 \zeta^\alpha_a\overline{\zeta}_{a\alpha}&=r^2_a\geq 0 \qquad  
\qquad \qquad \qquad &(c).
\end{array}
\end{equation}
Eqs (26) and (27) tell us that the $R_a$'s $(R_a=r_a^2\geq 0)$ 
are the radial variables and the  $\eta_a^\alpha $'s  and    
$ \overline {\eta}_{a\alpha} $'s,  which satisfy       
 \begin{equation}
\label{26}
\eta^\alpha_a\overline{\eta}_{a\alpha}=1;\qquad\eta^\alpha_a{\eta}_{a\alpha}= 
\overline{\eta}^\alpha_a\overline{\eta}_{a\alpha}=0,
\end{equation}
parametrize the two spheres $ S^2_a$. The $r_a^2$ and $\eta_a^\alpha $ and 
$ \overline {\eta}_{a\alpha} $  are in one to one correspondance with the 
$r$ FI isovectors. In other words eqs (28) describe a collection of $r$ 
unit two spheres which together with the $r_a$ conical variables of $(R^3)^r$ 
give the $3r$ parameters of  the $r$ FI isovector couplings.
\section {Separation of the charges  of the $U(1)_G$ gauge  and the  
$U(1)_R$ symmetries }
Here we describe the separation of  the gauge and $U(1)_R$  charges of the 
hypermultiplet scalar moduli  $\varphi_j^+$ and $ \overline{\varphi}_j^+$. 
As we  mentioned earlier, this charge separation is the second step in our 
programme  of finding the zero energy states  of the classical $2d$ $  N=4$
supersymmetric $U(1)^r$ gauge theory. Recall that the first step  decribed 
in section 3 consists to interpret $SU(2)_R$ field  representations as harmonic 
functions on $S^2_R$, fact which allowed us  to put eqs (1) in a form similar 
to eqs(2) as shown on eqs (25). However the field variables of eqs( 25) still 
carry  both $U(1)_R$ and $U(1)^r$ charges; these make their  geometrical 
interpretations difficult and moreover do not allow to take advantage  
with their similarities with   the  $N=2$ supersymmetric  D-flatness eqs (2) 
in looking for the solutions. Motivated by these two featurres, we have been 
lead to look for a way of rewriting eqs (25) so that the known  results of $2d$ 
$  N=2$ linear sigma  models, including  the geometric interpretation, can be 
exploited. We  have found that this way may be achieved by introducing a 
parametrization of the hyperkahler moduli where the gauge charges and the 
$U(1)_R$  ones are separated but  $2d$ $N=4$ supersymmetry still preserved. 
Note in passing  that  the idea of separating composite  quantum numbers of 
fields is not a new idea; and  corresponds just to a special feature  of group  
representations theory. In  the physics literature, the separation of  the 
different charges  of quantum fields has been used  succesfully in various  
occasions in the past,
in particular in coset models of  $2d$ conformal field theory and in strongly 
correlated electrons models of low dimensional systems. One of the well known 
examples concerns  complex spinors which may be separated into a spinon and a 
holon. For a review see [52]. Thus our  major aim in  this section is to use this 
idea and try to  solve these eqs by introducing  the  method of factorization of  
the two  kinds of charges carried by the hyperKahler moduli. There are various ways 
one may follows  in order to perform the  separation of gauge and $U(1)_R$ charges 
carried by $\varphi_j^+$'s and $ \overline{\varphi}_j^+$'s. A quite general way 
which allows  to fulfill the four following requirements ((a), (b), (c) and  (d))  
is given by splitting $\varphi_j^+$'s and $ \overline{\varphi}_j^+$'s as  shown 
on eqs (29). These requirements are natural  and  may be stated as follows:\\
(a) The  splitting  should preserves supersymmetry, that is preserving  the 
eight supercharges of the gauge theory. \\
(b)It should recover the results of [31,42 ] summarized in section 2 but also 
extend  the ADE models of $2d$ $  N=2$ supersymmetric backgrounds [8], see also [53]. \\
(c) It should give the standard ADE results up on breaking half of the eight 
supercharges.\\
(d) It should has a geometrical interpretation.\\
The factorization we propose is given by: 

 \begin{equation}
\label{27}
\begin{array}{lcr}
\varphi^+_j=X_j\eta^+_j+\gamma Y_j \overline{\eta}^+_j\\
\overline{\varphi}^+_j=\overline{X}_j \overline{\eta}^+_j-\gamma \overline{Y}_j 
\overline{\eta}^+_j,
\end{array}
\end{equation}
where $\eta^+_j$ and $ \overline{\eta}^+_j$ are as in eqs (27); that is:
\begin{equation}
\label{28}
\begin{array}{lcr}
\eta^+_j=u^+_\alpha\eta^\alpha_j;\qquad \overline{\eta}^+_j=u^+_\alpha
\overline{\eta}^\alpha_j \\
\eta^\alpha_j \overline{\eta}_{\alpha j}=1;\qquad \eta^\alpha_j {\eta}_{\alpha j}=
\overline{\eta}^\alpha_j \overline{\eta}_{\alpha j}=0 
\end{array}
\end{equation} 
and where $X_j$  and $Y_j$, $j=1,...,n$ are complex fields carrying no $U(1)_R$ 
charges. The parameter $\gamma$  takes the values $\gamma=0$  or $\gamma=1$  and 
distinguish the two classes of solutions we will give hereafter. Note that similar 
decompositions to eqs (29) are also valid for $\varphi_j^-$ and 
$ \overline{\varphi}_j^-$ and may be obtained from eqs (29) by acting  
on them by $D^{--}$ as in eqs (23) . We will not  use them in this 
discussion and then one can ignore them for the moment. Moreover the 
quantities $X_j$ , $Y_j$  and $\eta^+_j$ and $ \overline{\eta}^+_j$ of 
the splitting (29) behave under $ U(1)^r$ gauge and $ U(1)_R$ transformations 
as follows:        
\begin{equation}
\label{29}
\begin{array}{lcr}
U(1)^r : X_j \longrightarrow X^\prime_j=\lambda ^{q^j_a}X_j \\
\qquad Y_j\longrightarrow Y^\prime_j=\lambda ^{q^j_a}y_j \\
\qquad\eta^+_j\longrightarrow  \eta\prime^+_j= \eta^+_j  \\
\qquad \overline {\eta}^+_j\longrightarrow  \overline{\eta}\prime^+_j= 
\overline{\eta}^+_j ;
\end{array}
\end{equation}
and
\begin{equation}
\label{30}
\begin{array}{lcr}
 U(1)_R : X_j \longrightarrow X^\prime_j=X_j\\
\qquad Y_j\longrightarrow Y^\prime_j=Y_j\\
 \qquad\eta^+_j\longrightarrow  \eta\prime^+_j= e^{i\theta}\eta^+_j\\ 
\qquad \overline {\eta}^+_j\longrightarrow  \overline{\eta}\prime^+_j= 
e^{i\theta}\overline{\eta}^+_j .
\end{array}
\end{equation} 
Actually eqs (31-32) define the factorization  of  the gauge charges and 
$ U(1)_R$ ones. In what follows we shall use  the splitting (29) to solve 
eqs (1) which, by help of the analysis of section 3, is also given by         
\begin{equation}
\label{31}
\sum_{j}q^j_a \varphi_{j}^{+}\overline{\varphi}_{j}^{+}=r^2_a\eta_a^+
\overline{\eta}^+_a.
\end{equation}
We shall give two classes of solutions of these eqs; the first class described 
a generalisation of the usual  $N=2$ ADE ALE surfaces  and the second class 
describes new models which flow in the infrared to $2d$ $ N=(4,4)$ conformal 
field theories; the latters satisfy eqs (6). 
\subsection{ Generalized ADE hypersurfaces } 
Here we would like to  use the similarity betwen eqs (1) and eqs (2) in order 
to look for special solutions of eqs (1). These solutions are expected to   
describe  generalisations of the standard eqs of ADE singularities associated 
with $2d$ $N=4$ linear sigma models. We find indeed that the moduli space of 
gauge invariant vacua of eqs (33) is  appropriatly formulated in terms of harmonic 
variables of $SU(2)_R$ symmetry and the hypermultiplets vacua; see for instance 
eqs (35-36) for the case of a $ SU(n)$ singularity. As a check consistency of  
of our results, we show that under some assumptions to be specified later on,  
the generalized ADE hypersurfaces we have obtained may be brought to the well 
known ADE models of $ 2d$ $  N=2$ supersymmetric linear models. We show  also, 
by explicit computation, that for the case of a  $SU(n)$ singularity, the moduli 
space of gauge invariant vacua of eqs (33) is given by the usual ALE space with a 
$SU(n)$ singularity  times the 2-sphere to power $2n$; i.e ${(S^2)}^{2n}$. Under 
the abovementioned assymption, this reduces to the usual ALE background times a 
2-sphere.  A similar result is also valid for the other singularities. To do so, 
let us first note that up on imposing the condition of ADE models
\begin{equation}
\label{32}
\begin{array}{lcr}
\sum_{j}q^j_a=0,\qquad a=1,...,n-1,
\end{array}
\end{equation}
one can imitate the analysis of $ 2d $ $ N=2$  linear $ \sigma$ models 
and build the gauge invariant moduli in terms of the $\varphi_{j}^{+}$  fields. 
In the $SU(n)$ case for instance where $q_a^j$ is given by eq(3); there are three  
gauge invariant moduli; $U^{+{n(n+1)\over2}}$ , $V^{+{n(n+1)\over2}}$  
and $Z^{+{(n+1)}}$   carrying ${n(n+1)\over2}$ , ${n(n+1)\over2}$ and 
${(n+1)}$ $U(1)_R$  Cartan charges respectively. They are given by:      
\begin{equation}
\label{33}
\begin{array}{lcr}
U^{+{n(n+1)\over2}}=\prod\limits_{j=0}^n(\varphi^+_j)^{n-j}\\

V^{+{n(n+1)\over2}}=\prod\limits_{j=0}^n(\varphi^+_j)^j\\

Z^{+{(n+1)}}=\prod\limits_{j=0}^n(\varphi^+_j).
\end{array}
\end{equation}
They satisfy the following remarkable equation
\begin{equation}
\label{34}
U^{+{n(n+1)\over2}}V^{+{n(n+1)\over2}}=[Z^{+{(n+1)}}]^n
\end{equation}
Eq (36) generalizes the usual equation of the ALE surface with $SU(n)$ singularity 
which, for later use, we recall it herebelow:    
\begin{equation}
\label{35}
uv=z^n.
\end{equation}
 To better see the structure of eq (36), we use the splitting method of the  
charges of the $\varphi^+_j$'s we have described earlier. Taking $\gamma=0$, 
the general splitting eqs (29) reduces to: 
\begin{equation}
\label{36}
\varphi^+_j=x_j\eta^+_j ; \qquad \overline{\varphi}^+_j= \overline{x}_j 
\overline{\eta}^+_j,
\end{equation}
where $X_j$ and $\eta^+_j$ behave under gauge and $U(1)_R$ transformations 
as in eqs (31,32). Note by the way that like $\varphi_j^{\pm}$, the 
realization $X _j \eta_j^{\pm}$ carries , for each value of j, four real 
degrees of freedom; two degrees come from $ X_j$ and the two  others from  
the parameters  the 2- sphere  described by $\eta_j^{\pm}$; eqs (30). Under 
$2d$ $N=4$ supersymmetric transformations which  may be conveniently  
expressed as $4d$ $N=2$ supersymmetric transformations of fermionic parameters  
 $\epsilon ^{\pm}$  and $ \overline{\epsilon} ^{\pm}$, we have:
\begin{equation}
\label{38}
\delta\varphi^+_j=\epsilon ^+\psi_j+\overline{\epsilon} ^+ \overline{\chi}_j;
\end{equation}
where $\psi_j$ and $\overline{\chi}_j$ are  the Fermi partners of the 
$\varphi_j^{\pm}$ scalars. $(\varphi^{\pm}_j,\psi_j,\overline{\chi}_j)$ 
constitute altogether  the $ 4d$ $N=2$ free hypermultiplets. Using the 
splitting principle by factorizing  $ \epsilon^+$ as $\epsilon\eta^+$  
and $\overline{\epsilon}^+={\overline{\epsilon} } {\overline{\eta}^+}$, 
and using eqs (38) and (39); we get
\begin{equation}
\label{39}
\eta^+_j\delta X_j+X_j\delta\eta^+_j =\overline{\eta}^+\epsilon 
\psi_j+\overline{\eta}^+{\overline{\epsilon }} \overline{\chi}_j,
\end{equation}
or equivalenty
\begin{equation}
\label{38}
\eta^\alpha_j\delta X_j+X_j\delta\eta^\alpha_j =\overline{\eta}^\alpha\epsilon 
\psi_j+\overline{\eta}^\alpha\overline{\epsilon } \overline{\chi}_j.
\end{equation}
    Putting eqs (38) back into eqs (33), we get 
\begin{equation}
\label{37}
\sum_{j}q^j_a |X_{j}|^2\eta^+_j\overline{\eta}^+_j=r^2_a\eta_a^+\overline{\eta}^+_a
\end{equation}
and   
\begin{equation}
\label{38}
\begin{array}{lcr}
U^{+{n(n+1)\over2}}=uM^{+{n(n+1)\over2}}\\
V^{+{n(n+1)\over2}}=vN^{+{n(n+1)\over2}}\\
Z^{+{(n+1)}}=zS^{+{(n+1)}} ;
\end{array}
\end{equation}
where $ u, v, z$ and $M^{+{n(n+1)\over2}}$, $ N^{+{n(n+1)\over2}}$ and $ S^{+{(n+1)}}$ 
are gauge invariants given by
\begin{equation}
\label{39}
\begin{array}{lcr}
u=\prod\limits_{j=0}^n X_j^j&;\qquad N^{+{n(n+1)\over2}}&=
\prod\limits_{j=0}^n(\eta^+_j)^{n-j}\\
v=\prod\limits_{j=0}^n X_j^{n-j}&;\qquad M^{+{n(n+1)\over2}}&=
\prod\limits_{j=0}^n(\eta^+_j)^j\\
z=\prod\limits_{j=0}^n X_j&;\qquad S^{+{(n+1)}}&=\prod\limits_{j=0}^n\eta^+_j.
\end{array}
\end{equation}  
Note that  $ u, v$ and $z$ verify the relation (37) and $ M^{{+{n(n+1)\over2}}}$ , 
$ N^{{+{n(n+1)\over2}}}$ and $ S^{+{(n+1)}}$
satisfy eq (36). Eqs (42,43) may be brought to more familiar forms if we require 
moreover that the three following types of 2-spheres are identified:\\
(i) The $(n+1)$ 2- spheres parametrized by  the $\eta_j$'s .\\
(ii) The $(n-1)$ $\eta_a$ 2-spheres used in the parametrization of the FI couplings 
eqs (28). \\
(iii) The $ \eta^+ $ 2-sphere involved  in the factorization  of the supersymmetric 
parameter $\epsilon^+$ eq(40).\\
 In other words, we require the following identity: 
\begin{equation}\label{40}
\eta^+_j=\eta^+_a=\eta^+.
\end{equation}
With this identification eqs, (42) reduce to the well known 
D- flatness conditions of the $ U(1)_R$ gauge theory with four 
supercharges; namely:
\begin{equation}\label{41}
\sum_{j}q^j_a |X_{j}|^2=r^2_a.
\end{equation}
Moreover eq (36) reduce to the usual ALE surface with $SU(n)$ singularity 
eq (37) since the
 $M^{+{n(n+1)\over2}}$, $ N^{+{n(n+1)\over2}}$ and $ S^{+{(n+1)}}$ gauge 
invariant become trivial as they are given by powers of $\eta^+$ as shown here below.
 \begin{equation}\label{42}
\begin{array}{lcr}
M^{+{n(n+1)\over2}}=(\eta^+)^{{n(n+1)\over2}}=N^{+{n(n+1)\over2}}\\
S^{+(n+1)}=(\eta^+)^{n+1}.
\end{array}
\end{equation}
In the  general case where the gauge charges $q^j_a$ of the $X_j$'s satisfy the 
consraints (4), eqs (46) is the vaccum energy of $ 2d$ $ N=2$ supersymmetric 
linear  $sigma$ models. Thus the  classical moduli space M of the gauge invariant 
vacua of eqs (45,46) is then given by the 2-sphere parametrized by $\eta^+$; eq (45), 
times the moduli space of the gauge invariant solutions of $2d$ $ N=2$ supersymmetric 
vaccum energy states. In other words: 
$$ M={C^{n+1}\over {{C^*}^{n-1}}}\times S^2.$$
Note that  the identification constraint eq (45) has a nice interpretation; it  
breaks explicitly half of the eight supersymmetries leaving then four supercharges 
preserved. These  four supercharges are behind the reduction of eqs (42) down to eqs (46) leading to the standard ADE models. This  feature is immediatly derived by combining eqs (41)  and (45) 
as follows :
\begin{equation}
\label{39}
\eta^\alpha\delta X_j+X_j\delta\eta^\alpha =\overline{\eta}^\alpha\epsilon \psi_j+
\overline{\eta}^\alpha\overline{\epsilon } \overline{\chi}_j.
\end{equation} 
Then multiplying both sides of this identity by $\overline{\eta}_\alpha$; one gets, 
after using eqs (30): 
\begin{equation}
\label{39}
\delta X_j=\epsilon \psi_j, 
\end{equation}
giving the usual supersymmetric transformations of the complex scalars of the $2d$ 
$ N=2$ chiral multiplets. This completes the check of consistency of the generalised 
$ SU(n)$ hypersurface singularity (36). Before going ahead let us summarize in few  
words what we have done until now. Starting from eqs (1) , we have shown that it is 
possible to put  them into their equivalent form (33). The  corresponding moduli 
space of gauge invaraint vacua is given by eq (36) which reduces to the standard 
ALE space with $ A_{n-1}$ singularity up on imposing the factorization eqs (38) 
and  the conditions (45). The later breaks four supercharges among  the original 
eight ones. The factorization (38) offers in turns a method for a  geometric 
representation of hyperKahler backgrounds with eight supercharges.  
However we have not succeeded  to solve directly eqs (33) nor  (42) 
without breaking  the eight supercharges. We will see later that it  
sitll possible to work out solutions with eight supercharges by using 
the general splitting (29) instead of the factorization (38)  but still 
imposing eqs (45). Indeed to restore the eight supersymmetries by still using 
the constraint (45) we should take $\gamma$ non zero; say $\gamma=1$.  Non 
zero $\gamma$ brings four extra supercharges which add to the old  four existing 
ones carried by eq (38). This is easily  seen from the above analysis and the 
splitting (38) where each part  of the two terms of the right hand of eqs (39-41) 
carries four supersymmetries. We shall return to this feature  with more details 
in the next subsection; for the time being we would like to make two comments 
regarding eqs (33). \\
(1) A naive analysis  of eqs (33) suggests that the gauge invariant  moduli 
space of vacua  $M$ of eq (38) is given, for the generalized $SU(n)$ singularity 
eq (36), by  the usual ALE space with $SU(n)$ singularity times $2n$ two-spheres. 
In other words: 
$$ M={C^{n+1}\over {{C^*}^{n-1}}}\times{( S^2)}^{2n},$$
where  $(n+1)$ two spheres come from  the $ \varphi^+_j$'s as shown in eqs (38) and  
$(n-1)$ two spheres come from the FI couplings. \\
(2)  As far eq (36) is concerned, one can also write down the generalized ADE models 
extending the usual $N=2$ ones . In addition to eq (36) which generalizes eq (37) , 
we have also
$$ (x^{++})^{n}+x^{++}(y^{+(n-1)})^2+(z^{+n})^{2}=0,$$ describing  the generalized  
$D_n$ singularity extending the standard ALE one namely:
  $$ x^{n}+xy^2+z^{2}=0.$$
More generally, we have the following results:  
 $$(x^{+6})^2+(y^{+4})^3+(z^{+3})^4=0$$
$$(x^{+9})^2+(y^{+6})^3+y^{64}(z^{+4})^3=0$$
$$ (x^{+15})^2+(y^{+10})^3+(z^{+6})^5=0.$$
These eqs  extend respectively the following exceptional singulaerities 
$$ E_6:\qquad x^{2}+y^3+z^{4}=0$$
$$ E_7:\qquad x^{2}+y^3+yz^{3}=0$$
$$ E_8:\qquad x^{2}+y^3+z^{5}=0.$$ 
More informations about these extensions will be given in a future occasion.     
\subsection {Solutions with $\gamma=1$}
Choosing $\gamma=1$ in the eqs (29) and putting back into eqs (33), we get a 
system of three eqs given by:
\begin{equation}\label{43}
\begin{array}{lcr}
\sum\limits_{j}q^j_a(|X_j|^2-|Y_j|^2)\eta_j^+\overline{\eta}_j^+&=r^2_a\eta_a^+
\overline{\eta}_a^+ \qquad &(a)\\
\sum\limits_{j}q^j_a(X_j\overline{Y}_j)\eta_j^+{\eta}_j^+&=0 \qquad &(b)\\
\sum\limits_{j}q^j_a(\overline {X}_j{Y_j})\overline{\eta}_j^+\overline{\eta}_j^+&=0. 
\qquad &(c)
\end{array}
\end{equation}
At this level no constraint has been imposed yet  on the FI couplings contrary to
 the analysis of ref [31,42] summarized in section 2. If moreover we require that 
all the two sphere $\eta^+_j$ and $\eta^+_a$ are identified as in eq (45); the 
above system reduces to
\begin{equation}\label{44}
\begin{array}{lcr}
\sum\limits_{j}q^j_a(|X_j|^2-|Y_j|^2)&=r^2_a \qquad &(a)\\
\sum_{j}q^j_a(X_j\overline{Y}_j)&=0 \qquad &(b)\\
\sum_{j}q^j_a(\overline {X}_j{Y_j})&=0. \qquad &(c)
\end{array}
\end{equation}
Eqs (51) have some remarkable features which have nice interpretations. Though the 
$ q^j_a$ gauge charges of the hypermultiplets  are  not required to add to zero as 
in eq (4), eq (51.a) behave exactly as the D-flatness condition of $2d $ $N=2$ 
supersymmetric $U(1)^r$  gauge theory. The point is that eqs (51.a) involve twice 
the number of fields of eqs (2), but  with opposite charges $q^j_a$ . Put 
differently; eq (51) involve two sets of fields ${X_j}$ and ${Y_j}$ of charge 
$q^j_a$ and ($-q^j_a$) respectively. The sum of gauge charges of the $X_j$'s 
and $Y_j$'s add automatically to zero even  though eq (4) is not fulfilled. 
Thus models with $\gamma=1$ flow in the IR to a $2d$ $ N=(4,4)$ superconformal 
models extending the  usual $2d$ $ N=(2,2)$ ADE ones since the identities  
\begin{equation}\label{45}
\sum_{j}q^j_a+\sum_{j}(-q^j_a)=0
\end{equation}
go beyond the constraint eqs (4). Moreover 
eqs (51)  may be fulfilled in different ways;
either  by taking all charges $q^j_a$ of  the $U(1)^r$ gauge theory to be 
positive; say $q^j_a=1; a=1, ...,r; j=1,..., n$, or part of the $q^j_a$'s 
are positive and the remaining ones are negative. In the case of a $ U(1)$ 
gauge theory with $(n+1)$
hypermutiplets with gauge charges equal to one, eqs (51.a) describe a $ CP^n$ 
manifold whereas eqs (51.b-c) which read as
\begin{equation}
\label{46}  
 \sum_{j} X_j\overline {Y_j}=0 ,
\end{equation} 
together with their complex  conjugate, show that the  $\overline {Y_j},$'s are 
in the cotanget space of $ CP^n$ at the point $x_j=X_j/{[\sum\limits_i|Y_i|^2+r_a^2]^{-{1\over2}}}$.
Observe in  passing that in case where some of the positive charges $ q^j_a$ of 
$ U(1)^r$ gauge theory are not  equal to one, the corresponding moduli space is 
just  the cotangent bundle of some weighted  complex projective space, $ T^*(WP^n)$. 
Obesrve moreover that in the infrared limit this gauge theory flows to a $2d$  
$ N=(4,4)$ conformal field theory with central charge $C=6n$. In the next section 
we shall  give some illustrating examples. 

\section{Moduli space of vacua of models with $\gamma=1$}
\qquad In this section we want to study two types of vacua of the  D-flatness 
conditions of $2d $ $N=4$ supersymmetric $ U(1)^r$ gauge theory depending on 
the manner we deal with eqs(52). In other words starting from eqs(52),we develop 
hereafter two different, but equivalent, ways to solve them  . These ways are 
associated with the value of the sum over the $ U(1)^r $ charges of the hypermultiplet 
moduli that is ; ${ \sum \limits_{i}} q^i_a\ne0$ or ${ \sum\limits_{i}}q^i_a= 0$. 
To do so, we shall first study the case ${ \sum \limits_{i}} q^i_a\ne0$. We start 
by describing explicitly two examples after what we give the general sigma model 
result we have obtained and give also comments regarding the $2d$ $ N=4$ Liouville 
description in the viccinity of these singularities. A similar analysis will be 
made for the other case ${ \sum\limits_{i}}q^i_a= 0$.
\subsection{$ {\sum\limits_{i}}q^i_a\ne 0$}
A priori there are many ways to choose the $q^i_a$ charges such that 
$ \sum\limits_{i}q^i_a\ne 0$; each of which corresponds to a definite model. 
A simple and instructif model is to consider a $2d$ $ N=4$  supersymmetric 
abelian gauge theory with $(r+1)$
 hypermultiplets whose scalar fields are denoted as $\varphi^+_j$  and 
$ \overline{\varphi}^+_j;j=0,1,...,r$. Using  the splitting  method 
described previously, we write the $\varphi^+_j$'s  and $ \overline{\varphi}^+_j$'s as
\begin{equation}\label{47}
\begin{array}{lcr}
\varphi^+_j=X_j\eta^+ +Y_j \overline{\eta}^+\\
\overline {\varphi}^+_j=-\overline{Y}_j\eta^+ +\overline{X_j} \overline{\eta}^+,
\end{array}
\end{equation}
 where their $U(1)^r$
 charges are choosen as 
\begin{equation}\label{48}
q^j_i=\delta^j_{a-1}+\delta^j_{a}.
\end{equation}
Putting eqs (54) back into the D-flatness conditions (51) we get the following 
system of algebraic eqs 
\begin{equation}\label{49}
\begin{array}{lcr}
|X_{a-1}|^2+|X_a|^2-(|Y_{a-1}|^2+|Y_a|^2)= R_a \\
\sum\limits_j q^j_a X_j \overline{Y}_j=X_{a-1} \overline{Y}_{a-1}+X_a \overline{Y}_a=0\\
\sum\limits_j q^j_a \overline{X}_j {Y}_j=\overline{X}_{a-1} {Y}_{a-1}+
\overline{X}_a {Y}_a=0.
\end{array}
\end{equation}
For later use , let us rewrite the two leading blocks  of eqs of  the 
above system, describing respectively  models with $U(1)$ and $U(1)^2$ 
gauge groups associated with the   values  $r=1$ and $r=2$. For $ r=1$, 
eqs  (56) reduce to the three following eqs:
\begin{equation}
\label{50}
\begin{array}{lcr}
|X_0|^2+|X_1|^2-(|Y_0|^2+|Y_1|^2)= R_1&(a)\\
 X_0 \overline{Y}_0+X_1 \overline{Y}_1=0&(b)\\
\overline{X}_0 {Y}_0+\overline{X}_1 {Y}_1=0.&(c)
\end{array}
\end{equation}
Similarly we have, for  the $ U(1)^2$ gauge model, a system of six equations; 
three of them coincide with those given by eqs (57); the others are as follows:
\begin{equation}\label{51}
\begin{array}{lcr}
|X_1|^2+|X_2|^2-(|Y_1|^2+|Y_2|^2)= R_2&(a)\\
 X_1 \overline{Y}_1+X_2 \overline{Y}_2=0&(b)\\
\overline{X}_1 {Y}_1+\overline{X}_2 {Y}_2=0&(c)
\end{array}
\end{equation}
To solve eqs (56), we shall adopt the following strategy. We shall  first 
consider the solving  of eqs (57), then  we treat both eqs (57) and (58), 
after what we give the  general solutions for eqs (56) and  finally make some 
 comments regarding the Liouville description of the singularities of the metric 
of the Higgs branch. For the $ 2d$ $ N=4$ supersymmetric model with one $ U(1)$ 
gauge factor, one should note first of all that the moduli space of gauge 
invariant vacua is a complex surface which becomes singular when $R_1$ vanishes. 
It is just the cotangent line bundle of the two- sphere $S^2$; $T^*(CP^1)$. 
A naive way to see this feature is to set $ Y_0=Y_1=Y$; a choice which reduces 
eq (57.a) to the  following well known eq of $N=2$ linear sigma models with four 
supercharges 
\begin{equation}\label{52}
|X_0|^2+|X_1|^2-2|Y|^2=R.
\end{equation}
This eq describes the blow up of the $ SU(2)$ singularity of the ALE 
complex surface $ {C^2\over{Z_2}}$. An other way  to deal with this 
singularity is to make the change of variables preserving the eight 
supercharges  
\begin{equation}\label{53}
\begin{array}{lcr}
x_0=X_0[R_1+|Y_0|^2+|Y_1|^2]^{-{1\over2}}\\
x_1=X_1[R_1+|Y_0|^2+|Y_1|^2]^{-{1\over2}},
\end{array}
\end{equation}
 leading to  
\begin{equation}\label{54}
\begin{array}{lcr}
|x_0|^2+| y_1|^2=1&(a)\\
x_0 \overline{y}_0+x_1 \overline{y}_1=0&(b)\\
x_0 {y}_0+\overline{x}_1 {y}_1=0&(c).
\end{array}
\end{equation}
 Eqs (57.b-c), which by the way, are exchanged under complex conjugation, 
have a geometric meaning; they show that at  each point  $(x_0, x_1)$  of 
the base manifold $B_1$ there is an orthogonal fiber $F_1$ parametrized by 
$(\overline{Y}_0, \overline{Y}_1)$, defining altogether the cotangent line 
bundle $T^*CP^1$. For non zero values of $ R_1$ where the change (60) is well 
defined, eq (57.a) is a 2-sphere and then the bundle is smooth. For $ R_1$ 
equals to zero, the change (60) falls down at the origin $ X_0=X_1=Y_0=Y_1=0$ 
and the bundle becomes singular.  Note  that according to the ADHM construction, 
the moduli space of gauge invariant vacua of the $U(1)$ gauge model with two 
(or more) hypermultiplets is just the moduli space of small instantons on $R^4$. 
For $R_1$ positive definite, the small instanton singularity is blown up and 
in the limit  $R_1=0$ the singularity is recovered. In two dimensions,
it has been shown  moreover that in this limit the fields $X_0$, $X_1$, $Y_0$ and $Y_1$  do not give a good description  of the small instanton  conformal theory near the singularity. The appropriate variables in this region turns out to be  those of  a $2d$ $N=4$ conformal Liouville  field theory [17,18 ]. To see this remarkable feature, it is interesting to use the field theoretical approach of Aharony and Berkooz [18 ] regarding  the study of the low energy lmits of $ 2d$ $N=4$ gauge theories whose lagrangian  $ L=L_{gauge}+L_H$  reads as:
\begin{equation}\label{54}\begin{array}{lcr}
L={1\over{4g^2_{YM}}}\int d^2x tr (F^2_{\mu\nu}+(D_\mu V)+[V,V]^2+ 
\overline{\psi}_V \gamma^\mu D_\mu \psi_V+ \overline{\psi}_V [V,\psi_V]+
\vec{D}^2)+\\ \int d^2x \sum\limits _{hypermult} (|D_\mu \varphi_H|^2+
|V\varphi_H|^2+ \overline{\psi}_H \gamma^\mu D_\mu \psi_H+\overline{\psi}_H V{\psi}_H+
\overline{\psi}_VV{\psi}_H \\+ \overline\varphi_H D\varphi_H)
\end{array}
\end{equation}
In this formal eq $ D_\mu=(\partial_\mu+A_\mu)$ is the covariant derivative, 
$(V_{A \overline {A}}, A_\mu, D^{(\alpha\beta)})$ and $(\psi^A_V, 
\psi^{\overline{A}}_V)$ are respectively the bosonic and fermionic 
fields of the vector multiplet carrying amongst others quantum charges 
of the $ SO(4)\times SU(2)_R\approx SU(2)_r \times SU(2)_l \times SU(2)_R$ . 
Note that $SU(2)_r \times SU(2)_l \times SU(2)_R$ is the R-symmetry of the
 gauge theory with eight supercharge in two dimensions. Note also that the 
indices $A$,$ \overline{A}$ and $\alpha$  refer to the isospin $1\over2$ 
representation of $SU(2)_l , SU(2)_r$ and $SU(2)_R$ respectively. Note Moreover that 
  the $\varphi_H$ scalars and their fermionic partners $(\psi^L_H, \psi^R_H)$ stand 
for the fields of the hypermutiplets. Following [18], the low energy limit of this 
gauge theory, which involves taking $ g_{YM} \to \infty$, is described by two
 decoupled $2d$ $N=(4,4)$ superconformal field theories; one describing the Higgs
 branch whose central charge $C_H=6(n_H-n_V)$  where  $n_H$  the number of
 hypermultiplets and $n_V$ is the number of vector multiplets. The other 
conformal field theory  corresponds to the Coulomb Branch of central charge
 $C_V=6$. An argument supporting this paticular feature comes from the analysis 
of the R-symmetries of  the $ N=(4,4)$  superconformal algebra which includes 
left and right moving $su(2)$ Kac Moody subalgebras. The R-symmetry of the 
Higgs branch is exactly $ SU(2)_l \times SU(2)_r\approx SO(4)$ encountered 
earlier while the R-symmetry of the  Coulomb branch is given by a non visible 
group $SO(4)\approx SU(2)_l \times SU(2)_r$ containing $SU(2)_R$ as a diagonal 
subgroup. Since the Coulomb  and Higgs  superconformal theories have different 
R-symmetries; they cannot be identified. Moreover taking  the naive limit 
$ g_{YM} \to \infty$ in eq (62), one sees  $L_{gauge}$  is removed and  
the lagrangian of  the low energy gauge theory  is reduced to $L_H$; the 
lagrangian of the Higgs branch namely: 
\begin{equation}
\label{equation}
\begin{array}{lcr}
L_H=\int d^2x \sum\limits _{hypermult} [|D_\mu \varphi_H|^2+|V\varphi_H|^2+ 
\overline{\psi}_H \gamma^\mu D_\mu \psi_H+\overline{\psi}_H V{\psi}_H+
\overline{\psi}_VV{\psi}_H \\+ \overline\varphi_H D\varphi_H]
\end{array}
\end{equation}
where now the vector  multiplet fields are auxiliary fields which may be 
eliminated through their eqs of motion. However following [18] see also 
[54], it more useful to regard the vector multiplet fields as the basic 
objects instead of the matter fields and integrating  over the hypermultiplet 
fields in order to describe the behavior near the singularity in the moduli 
space. In this lagrangian approach, one obtains an induced effective action of 
the vector mutiplet fields which describe the region near the singularity of the 
Higgs branch $(\varphi\to 0 $ or $ V\to\infty)$. In the case of supersymmetric 
$U(1)$ gauge theory with $ N_f$ hypermultiplets and one vector multtiplet , 
supersymmetry and $SO(4)$
symmetry constraint the metric of the four gauge scalar fields 
$(V_i)=(V_1,V_2,V_3,V_4)$ in the vector multiplet to be of the form:
\begin{equation}\label{55}
ds^2=N_f{1\over{[(V_1)^2+V_2^2+V_3^2+V_4^2]}}[({dV_1}^2)+{(dV_2}^2)+{(dV_3}^2)+{(dV_4}^2)].
\end{equation}
or equivalently by changing to radial cordinates $\sum\limits_{m=1}^4 (dV_m)^2=
 {dv}^2+ v^2\sum\limits_{i=1}^3 (d\Omega_i)^2$, and defining a new variable 
$ \phi= {\sqrt{N_f\over 2}}\log ({v\over M})$  for some mass scale $M$:
\begin{equation}\label{55}
ds^2=d\phi^2+{N_f\over2}\sum\limits^3_{i=1}(d\Omega_i)^2,
\end{equation}
 together with the 3-form torsion $H$ given  by $(-N_f)$ times the volume form 
of the 3-sphere namely: 
\begin{equation}\label{55}
H=-N_fd\Omega_1d\Omega_2d\Omega_3=-N_f\bf{d\Omega}
\end{equation} 
 The effective theory in the region of large $V$ is described by a Liouville 
field $\phi$, its fermionic partner  $\psi_\phi$ and a supersymmetric level 
$N_f$ $SU(2)$ WZW model  generated by the usual currents $J^{\pm}$ and  $J^3$ 
which  may be rewritten as the sum of a bosonic level $(N_f-2)$ $SU(2)$ WZW model
 plus three free fermions. $\psi^{\pm}_{SU(2)}$ and  $\psi^{3}_{SU(2)}$  Altogether
 these fields give a realization of the $N=4$ conformal field theory of the central 
charge $ C=6(N_f-1)$ as shown on the following central charge counting
\begin{equation}\label{55}
 6(N_f-1)=2+{3(N_f-2)\over{N_f}}+(1+3Q^2);
 \end{equation} 
where $Q=(N_f-1)\sqrt{2\over{N_f}}$. In the end of this digression on  the physics 
in the throat of the Higgs branch, note that the Liouville field $\phi$ is 
intimately related with the four scalars $\{V_m\}$ of the vector multiplet and 
then with the abelian $U(1)$ gauge factor as shown on the following eq. 
\begin{equation}
\label{55}
\exp({\sqrt2\over{N_f}}\phi)\sim\sqrt{V_1^2+V_2^2+V_3^2+V_4^2}.
 \end{equation} 
Therefore there is one to one correspondance  between the liouville field $\phi$ 
the vector multiplet of the $2d$ $N=4$ $U(1)$ gauge theory In other words the 
field $\phi$ is one to one correspondance with the $U(1)$ factor of the gauge theory. 
In this regards, one ask the following question. What happens if, instead of $ 2d $
 $N=4$ supersymmetric $U(1)$ gauge theory , we consider a $U(1)^r$ gauge theory 
 involving $r$ abelian $U(1)$ factors and then $4r$ scalars $ V_{m,a}$; 
$a=1,..,r; m=1,2,3,4$? Before discussing the answer to this question, 
let us first consider the linear sigma model solutions for  a typical $U(1)^r$ 
D-flatness eqs. This concerns for example of type eqs (57-58) which  are associated 
with a $U(1)\times U(1)$ supersymmetric  linear sigma model. Following the same 
steps we described above, one can solve these eqs in a similar way as for the $ U(1)$
 theory. The result is that eqs (57-58) describe a two dimensional complex surface 
given by two intersecting smooth $T^*CP^1$'s of base 
 \begin{equation}\label{55}
\begin{array}{lcr}
|x_0|^2+|x_1|^2=1\\
|\overline{x}_1|^2+|\overline{x}_2|^2=1,
\end{array}
\end{equation}
where $|\overline{x}_1|$ and $|\overline{x}_2|$ are obtained from eqs (57-58) 
and analogous  changes as in eq (60). Using the results of [17,18] and the 
discussions made in the end of the previous example, one sees that here also 
the fields $X_i$ and $Y_i$ could not be the  appropriate variables in the 
viccinity of the  singularity.  Since this singularity is a degenerate 
singularity of type $A_2$, we expect to have more than one Liouville 
mode in this region and then a more general $2d N=4$ conformal field theory 
with backgound charges. A naive way to see this feature is to use the radial 
coordinates change of the $ U(1)$ gauge theory which allowed us to put eq(64)
 into its equivalent form eqs ( 65,66). Since in the $U(1)\times U(1)$ gauge 
theory we are discussing we have two kinds of scalar fields $ V_{1,m}$  and 
$ V_{2,m}$ corresponding to each $U(1)$ factor of the $U(1)\times U(1)$ group, 
one is tempted to extend the above radial charge to lead to a  $N=4$ conformal 
$su(3)$ Toda theory . Indeed, starting from the radial parametrization  
\begin{equation}\label{55}
\sum\limits_{m=1}^4|dV_{\rho,m}|^2=(dv_\rho)^2+{v_\rho}^2\sum\limits_{i=1}^3
(d\Omega_{\rho,i})^2; \rho=1,2
\end{equation}
 and introducing two  scalar fields  $\phi_\rho$:
\begin{equation}\label{55}
\phi_\rho=a_\rho\log {v_\rho \over M}
\end{equation}
where the coefficients $a_\rho$  should be determined by   
$2d$ $N=4$ conformal invariance;  one can write down an extension of 
eqs (65,66). Supersymmetry and $SO(4)$ invariance suggest the following  
extension :
 \begin{equation}\label{55}\begin{array}{lcr}
ds^2= {1\over2}K_{\rho\sigma}[d\phi_\rho d\phi_\sigma+a_\rho a_\sigma 
\sum\limits_{i=1}^3 d\Omega_{\rho,i} d\Omega_{\sigma,i}];\\
H=-2a_\rho d\Omega_\rho,
\end{array}
\end{equation}
where $K_{\rho\sigma}$ is $ su(3)$ the Cartan matrix . More generally,  
this analysis may be extended in a natural way to any  $2d$ $N=4$ $U(1)^r$ 
gauge theory $r\geq 1$ in presence of $N_{f,r}$ hypermultiplets. To do so one 
should first note that for a $2d N=4$ supersymmetric $U(1)^r$  linear $\sigma$ 
model with $(r+1)$  hypermultiplets the moduli space is given by the intersection 
of $r$ $T^*CP^1$'s. When all the FI coupling  variables vanish simultaneously, 
the physics within the  Higgs branch throat is expected to be described by a 
general $ 2d$ $N=4$ superconformal Toda theory. In this region the metric is 
expected  to have  a form like that given by eqs(65,66).  Progress in this 
direction will be reported elsewhere [55]. 
\subsection{$ \sum\limits_{i } q^i_a=0$}
 This situation is  the relevent one in the analysis of the moduli space of gauge 
invariant vacua of $ 2d N=2$ supersymmetric linear sigma models. It ensures that 
in the infrared, the  gauge theory flows to a  superconformal one and plays a 
crucial role in the study of superstrings compactifications on  local Calabi 
Yau manifolds with ADE singularities. The $ q^j_a$'s  satisfying the relation 
{$ \sum\limits_{i } q^i_a=0$} are also one of the main ingredient in toric geometry 
especially  in the toric construction of Calabi Yau manifold and their mirrors [56].
\\ In the case of $2d$ $ N=4$ supersymmetric linear sigma models we have been studing, 
the sum over the $ q^j_a$ charges is automatically fulfilled as shown on eq (52) 
and then one maight conclude  that it is not  necessary  to distinguish the two 
senarios described in paragraphs 5.1 and 5.2. Though this remark is partially true, 
there are however some remarkable subtilities we will comment in a moment. Moreover,
 distinguishing the two senarios is also relevant for studying $N=4$ supersymmetric 
backgrounds by imetating  methods of $ 2d$ $ N=2$  supersymmetric  linear models as 
we have done in subsection 4.2 . In what follows we give two examples illustrating 
the above remarks . In the first example we consider $ 2d$ $ N=4$ $ U(1)^2$ linear 
sigma model with three hypermultiplets of $ q^j_a$ charges chosen as: 
 \begin{equation}\label{56}
q^j_a=\delta^j_{a-1}-\delta^j_{a}; \qquad  \sum_{j=0}^{2}q^j_a=0.
\end{equation}
 The D-flatness conditions, which may be deduced from eqs (56) and ( 73 ) read as  
\begin{equation}
\label{57}
\begin{array}{lcr}
(|X_0|^2-|X_1|^2)- (|Y_0|^2-|Y_1|^2)=R_1\\
X_0\overline{Y}_0-X_1\overline{Y}_1=\overline{X}_0 Y_0-\overline{X}_1Y_1=0,
\end{array}
\end{equation}
together with
\begin{equation}
\label{58}
\begin{array}{lcr}
(|X_1|^2-|X_2|^2)- (|Y_1|^2-|Y_2|^2)=R_2\\
X_1\overline{Y}_1-X_2\overline{Y}_2=\overline{X}_1 Y_1-\overline{X}_2Y_2=0.
\end{array}
\end{equation}
 A way to handle these eqs is to note that they are quite similar  to eqs (57,58) 
up to permutating the roles of $X_1$, $X_2$ and $\overline{Y}_1$ and $\overline{Y}_2$ 
respectively. From this view point eqs (74 ) may be rewritten as
 \begin{equation}
\label{58}
\begin{array}{lcr}
(|X_0|^2+|-\overline{Y}_1|^2)- (|Y_0|^2+|X_1|^2)=R_1\\
X_0\overline{Y}_0+X_1(-\overline{Y}_2)=\overline{X}_0 Y_0+\overline{X}_1(-Y_2)=0.
\end{array}
\end{equation}
 For later use it is intersting to rename the feld variables  of eqs (76)as 
$X_0=Z_0$,$(-\overline{Y}_1)=Z_1$; $Y_0=W_0$ and  $X_1=W_1$. Putting this change 
in the above eqs, one sees that the resulting relations are comparable to those 
given by eqs (58,61 ). Thus eqs (76) describe just a cotangent bundle of $CP^1$. 
The base $B_1$ and the fiber $F_1$ are respectively parametrized  by the local 
coodinates  $(z_0,z_1)$ and $(w_0,w_1)$  where the $z_i$'s and $w_i$'s are related 
to $ Z_i$'s and $W_i$'s by analogous formulas to those given by eqs (60). In the 
case of the $ 2d$ $ N=4$ supersymmetric $ U(1)\times U(1)$ gauge theory, we have to
 solve the system of eqs (74) and (75) which we rewrite for convienience as 
\begin{equation}
\label{57}
\begin{array}{lcr}
(|Z_0|^2+|Z_1|^2)- (|W_0|^2+|W_1|^2)=R_1\\
Z_0W_0+Z_1W_1=\overline{Z}_0\overline{W}_0+\overline{Z}_1\overline{W}_1=0,
\end{array}
\end{equation}
 Eqs (77) coincide with  eqs (76) we have considered  above while eqs (75) read now as  
\begin{equation}
\label{57}
\begin{array}{lcr}
(|W_1|^2+|Z_2|^2)- (|Z_1|^2+|W_2|^2)=R_2\\
Z_1W_1+Z_2W_2=\overline{Z}_1\overline{W}_1+\overline{Z}_2\overline{W}_2=0,
\end{array}
\end{equation}
 where we have set $\overline {Y}_2=Z_2$ and $X_2=W_2$. For positive definite 
values of $ R_2$, if we take $Z_1=W_2=0$, one sees that the complex coordinates 
$(W_1, Z_2)$ parametrize a $CP^1$ complex curve which is isomorphism to a real 
$2$-sphere of radius $\sqrt{R_2}$. In the limit when $R_2$ goes to zero, this 
two sphere collapse and one ends with a $SU(2)$ singularity. For generic values 
of $Z_1$ and $W_2$, eqs (78) describe a cotangent bundle: $T^*CP^1$ exactly as 
for eqs (77), the radius of the base $B_1$ of $T^*CP^1$ is proportional to 
$\sqrt{R_1}$. Eqs (77) and (78) describe then two intersecting  cotangent 
$ CP^1$'s whose bases $B_1$ and  $B_2$ as well as fibers $F_1$ and $ F_2$ 
are roughly speaking parametrized by $(Z_0,Z_1)$,$(W_1,Z_2)$,  $(W_0,Z_1)$ 
and $(W_2,Z_1)$ respectively. The second example we want to give deals with 
the case of a $2d$ $N=4$ supersymmetric $U(1)^r$ gauge theory with $ r+2$ 
hypermultiplets $\varphi^+_j$ of charges $q^j_a$ given by
\begin{equation}
\label{3}
\ q_a^j=-2\delta_a^j+\delta_a^{j-1}+\delta_a^{j+1},
\end{equation}
 satisfying the identity
 \begin{equation}
\label{3}
\sum _{i=0}^{r+1}q_a^j=0.
\end{equation}
Using the splitting (29)  with  $\gamma=1$, one sees that the $X_j$'s  
and $Y_j$'s transform under the $ {C^*}^r $ actions in the same manner as  
the $\varphi^+_j$ namely:
\begin{equation}
\label{3}
\begin{array}{lcr}
X_j \longrightarrow\lambda ^{q^j_a}X_j \\

 Y_j\longrightarrow \lambda ^{q^j_a}Y_j ;
\end{array}
\end{equation} 
 where $\lambda$ is non zero complex parameter. Putting eqs (79) into the 
D-flatness eqs (56 ), one gets the following system of $3r$ eqs:
\begin{equation}
\label{58}
\begin{array}{lcr}
(|X_{a-1}|^2+|X_{a+1}|^2-2|X_a|^2)- (|Y_{a-1}|^2+|Y_{a+1}|^2-2|Y_a|^2)=R_a &(a)\\
X_{a-1}\overline{Y}_{a-1}+X_{a+1}\overline{Y}_{a+1}-2X_a\overline{Y}_a=0 &(b)\\
\overline{X}_{a-1}{Y}_{a-1}+\overline{X}_{a+1}{Y}_{a+1}-2\overline{X}_a{Y}_a=0.&(c)
\end{array}
\end{equation}
 For the simple example of the $U(1)$  gauge theory, one can check by 
following the same procedure we described in the previous example that 
the system of eqs given herebelow
\begin{equation}
\label{58}
\begin{array}{lcr}
(|X_{0}|^2+|X_{2}|^2+2|Y_1|^2)- (|Y_{0}|^2+|Y_{2}|^2+2|X_1|^2)=R_1\\
X_{0}\overline{Y}_{0}+X_{2}\overline{Y}_{2}-2X_1\overline{Y}_1=0\\
\overline{X}_{0}{Y}_{0}+\overline{X}_{2}{Y}_{2}-2\overline{X}_1{Y}_1=0,
\end{array}
\end{equation}
describe the cotangent bundle of the complex two dimensions weigthed  
complex pojective space $WP^2_{1,2,1}$. For the general $U(1)^r$ gauge 
theory, if all the $R_a$'s are non zero,  eqs (82) describe  the intersection 
of $r$ $WP^2_{1,2,1}$ cotangent bundles.\\
In the end of this discussion we would like to make a comment regading the 
second example. This concerns the link between our present analysis and the 
usual ADE $ N=2$ syupersymetric models. Starting from the splitting (29) of 
the hypermultilpet moduli, one may  view the $X_j$'s  and $Y_j$'s as  vacua 
of the moduli space of two orthogonal copies of $2d$ $N=2$ supersymmetric 
$A_r$ models. This feature is easily seen on the $ {C^*}^r $ action on these 
fields as shown on eqs (81) and may be rendered more manifest by analysing 
the  D-flatness eqs (82). Ignoring for the moment eqs (82.b-c) and setting 
$R_a=A_a-B_a$ with $A_a \geq B_a $, one may put eqs (82.a) into the following 
remarkable form describing two copies of $A_r $ models
\begin{equation}
\label{58}
\begin{array}{lcr}
(|X_{a-1}|^2+|X_{a+1}|^2-2|X_a|^2)=A_a\\
 (|Y_{a-1}|^2+|Y_{a+1}|^2-2|Y_a|^2)=B_a\\
\end{array}
\end{equation}
For $A_a=0$; $B_a$ positive definite  or $A_a$ positive definite; $B_a=0$, 
one of the $A_r$ models is singular while $A_a=B_a=0$ both of them are singular.
 For $A_a$ positive definite; $B_a$  positive definite, eqs( 84) describe the 
blown up of the two $A_r$ singularities. Note that each of the $ A_r$ models has 
$N=2$ supersymmetry whereas the origonal eq (82) from which they come have $N=4$ 
supersummetry. This means that eqs( 82.b-c) are the lacking piece that rotates 
the two orthogonal $N=2$ supersymmetries. Eqs (82.b-c) reflect the neceessary 
conditions to get $N=4$ supersymmetry in two dimensions  starting from two 
orthogonal blocks of $N=2$ supersymmetric models. From this naive parametization 
of the two intersecting $T^*CP^1$'s, one sees that the base $B_1$ intersects the 
fiber $F_2$ along the $ Z_1$ direction and the fiber $ F_1$ intersects the base 
$B_2$ along the $\omega_1$ direction.\\
\section{Conclusion}
 In this paper we have studied two main things. First, we have developed the 
analysis of the resolution of ADE singularities of hyperKahler manifolds involved
 in strings compactification. This concerns too particularly the moduli spaces of
 the Higgs branch of supersymmetric $U(1)^r$ gauge theories with eight supercharges. 
Second, we have initiated the analysis of singular CFT's with higher order degeneracies 
by using the field theoretical approach of Aharony and Berkooz. Actually this study may
 be viewed as an extension of the recent works dealing with the leading  $A_1$ 
singularity.\\
Concerning the first part, we have studied the solutions of the D-flatness 
eqs of supersymmetric $ U(1)^r$ gauge theories with eight supercharges by 
using the linear sigma model approach. We have given, amongst others, a 
geometrical interpretation  of the blown up singularities as a collection  
of intersecting cotangent complex dimensional weighted projective spaces 
depending on the number of hypermultiplets and gauge supermultiplets. This 
examination extends the standard linear sigma model analysis performed for 
the Kahler Coulomb branch of supersymmetric gauge theories  with four supercharges. 
Our way of doing go beyond literature analysis where only half of the eight 
supersymmetries are manifest. Our method preserves manifesty all the eight 
supersymmetries and is realised in two steps  based  on a geometric realization 
of the $SU(2)_R$  symmetry on one hand and on a separation of the charges of the 
gauge and $R$-symmetries on the other hand. The factorisation of the gauge and 
$SU(2)_R$ charges of the hypermultiplets moduli involves  a parameter $\gamma$  
taking the values $\gamma=0$ or $\gamma=1$ which distinguish two classes of 
solutions of eqs(1) both preserving the eight supercharges. For $\gamma=0$, 
we have obtained a generalisation of the ADE complex surfaces reproducing the 
standard ones by partial breaking of $2d$ $N=4$ supersymmetry down to $2d$ $N=2$. 
For $\gamma=1$, we have found new models which flow in the infrared to $2d$ $N=(4,4)$ 
scale invaraint models. In this context several examples are given and classified  
according the manner one solves eqs(6). In the second part of this paper, we have 
studied the infrared dynamics of two dimensional $N = (4,4)$ gauge theories using 
field theoretical methods. We have made comments regarding the $ N=4$  conformal 
Liouville description of the region in the viccinity of the singularity of the 
metric of the  $2d $ $N=4$ Higgs branch. In this region, the string coupling 
constant $g_s=e^\phi$ blows up as the Liouville field $\phi$ goes to infinity 
[17,18]. In an attempt towards an interpretation of the degenerate  $ A_r$ 
singularity  carried by eqs (1), we have given field theoretical arguments 
suggesting that the metric of the moduli space near the Higgs singularity  
maight be described  by a $N=4$ conformal $SU(r+1)$ Toda theory in two dimensions.
 This observation needs however a detailed  study. In this regards a project of 
checking this observation for the case of the $sp(2)$ gauge group is understudy 
[55].\\
 \bf {Aknowledgements}\\
One of us(EHS) would like to thank Profs J.Shinar and J.Vary for kind 
hospitality at IITAP, Iowa State University; where a part of this work is done. 
AB would like to thank the organizers of the Spring Workshop on Superstrings and 
related Matters (March 1999), The Abdus Salam International Centere for Theoretical 
Physics Trieste ,Italy, for hospitality.  \\

 This research work has been supported  by the program PARS number 372-98 CNR.\\
\bf References
\begin{enumerate}
\item [[1]]  N. Seiberg and E. Witten , Nucl. Phys (1994)B426-19,hep-th/9407087. 
 \item [[2]] N. Seiberg and E. Witten , Nucl. Phys (1994)B431-498,hep-th/9408099.
 \item [[3]] A. ELHassouni, T.Lhallabi, E.Oudrhiri and E.H.Saidi; Class.Quant.Grav 5 
(1988)287-298.
 \item [[4]] A. Hanany and E. Witten ; Nucl. Phys  B492 152-190 (1997),hep-th/9611230.
 \item [[5]]  E. Witten ; Nucl. Phys  B500 3-42 (1997),hep-th/9703166.
\item [[6]] A. Klemm, W. Lerche , P Mayr , C Vafa and N. Warner; Nucl Phys B477(19997) 
 \item [[7]] S. Katz,A.Klemm and C. Vafa; Nucl. Phys (1997) B497 173-195.   
\item [[8]] S. Katz, P. Mayr and C. Vafa, « Mirror symmetry and exact solution of 4d 
N=2 gauge theories I » Adv,Theor. Math.Phys(1998)1.53.
\item [[9]]  N.C. Leung and C. Vafa ; Adv .Theo . Math. Phys 2(1998) 91, hep-th/9711013   
\item [[10]] C. Vafa, Exact Results of N=2 Compactifications of Heterotic Strings, 
Summer School in High Energy  Physics and Cosmology  (ICTP, Trieste,1996).
\item [[11]] C. Vafa, «  Geometric Physics»; Talk presented at  the ICM' 98. 
 \item [[12]]  P. Mayr , Geometric Construction of N=2 gauge theories, Spring 
Workshop on Superstrings and Related Matters, (ICTP, Trieste, March 1999).  
\item [[13]] A.Belhaj,  A. EL Fallah and E.H. Saidi,« On the affine$ D_4$ mirror 
geometry;  Class. Quantum. Grav. 16 (1999)3297-3306.
\item [[14]] A.Belhaj,  A. EL Fallah and E.H. Saidi« On non simply laced  
mirror geometries in type II strings;   Class. Quantum. Grav. 17 (1999)1-18.
\item [[15]A. Shapere  and C. Vafa, «  BPS Structrure of ARgyress -Douglas 
Superconformal theories » ;hep-th/9910182. 
\item [[16]]   T. R. Taylor and C. Vafa, « RR flux on Calabi Yau and Partial 
Supersymmetry Breeking  » ;hep-th/9912152.
\item [[17]]N. Seiberg and E. Witten,«  D1-D5 and singular CFT»; JHEP 9904 (1999) 
017, hep-th/9903224
\item [[18]] Aharoni, Berkooz « IR Dynamics of d=2 , N=(4,4) Gauge Theories and DLCQ 
of Little String theories »; JHEP 9910 (1999) 0030, hep-th/9909101.
\item [[19]] A. Given , D. Kutasov and N. Seiberg ,«  More Comments on String on 
$ADS_3$  » ; hep-th/9806194.
\item [[20]] D. Kutasov and N. Seiberg, «   Comments on String on $ADS_3$  » ; 
Adv .Theo . Math. Phys 2(1998) 733,  hep-th/9806194.
\item [[21]]  E. Diakonosko and N. Seiberg ,«   The Coulomb branch of (4,4) 
supersymmetric Field theories in two dimensions », hep-th/9707158 ,JHEP 9707(1997)1.
\item [[22]] J. Brodie «Two Dimensions Mirror Symmetry from M-theory »; Nucl. Phys  
B517(1998)36-52, hep-th/9709228.
\item [[23]] S. Sehti,  «  The Matrix Formulation of type II Five branes »; Nucl. 
Phys  B52(1998) 158,hep-th/9710005. 
\item [[24]] K. Schoutens, « O(N) Extended Superconformal Field Theory In Superspace»; 
Nucl. Phys  B295(1998)634.
 \item [[25]] A. Servin, W. Troost and A. Van Prooeyen, «  Superconformal Algebras 
In Two dimensions with N=4 »;  Nucl. Phys B304(1988)447.
\item [[26]] E. A.Ivanov , S. O. Krivonos and V. M. Leviant  ,«  A New Class Of 
Superconformal Sigma models  With the Wess -Zumino action »; Nucl. Phys B304(1988)601. 
 \item [[27]] C. Kounass , M. Porrati and B. Rostand, «   On N=4 extend super-Liouville 
Theory »;  Physs. Lett B225 (1991)611.  
\item [[28]] E. H. Saidi and M. Zakkari, Class Quant Grav 7 (1991) 123; 
E. H. Saidi and M. Zakkari,  Inter Jour Mod A6 (1991) 3175.
\item [[29]] S. Kachru and  C. Vafa, «  Exact Results for N=2 compactifications 
of Heterotic Strings »; Nucl Phys B 450(1995) 69, hep-th/9505105. 
 \item [[30]] S. Ferarra, J.A. Harvey, A . Strominger, and C Vafa, «  Second -Quantized mirror Symmetry »; Physs. Lett B361 (1995)59-65.   hep-th/9505162. 
\item [[31]] E .Witten, «  Heteroric String Conformal Field Theory ad ADE 
Singularities », hep-th/9909299.
\item [[32]]N. Seiberg and E .Witten «   Gauge Dynamics and Compactification to 
Three Dimensions  » In The mathematique Beauty of Physics , hep-th/9609122 .
\item [[33]]K. Hori, H. Ooguri  and C. Vafa, « Non -Abelian  Conifold 
Trasitions and N=4 in Three dimensions » ; Nucl. Phys  B504 117 (1997), hep-th/9705220.
\item [[34]] Moshe Rozali, «  Hypermultiplet  Moduli Space and Three 
Dimensional Gauge Theories  »; hep-th/9910238 
\item [[35]] P. Mayr, «  Conformal Field Theory on $K3$ and Three dimensional gauge theories »; hep-th/9910268. 
\item [[36]]E. Witten, « String  Theory Dynamics in Various Dimensions »; 
Nucl Phys B 433(1995) 85, hep-th/9503124.
\item [[37]] P.S. Aspinwall, «  Enhanced Gauge Symmetries and $K3$  surfaces »;  Phys . Lett B443 (1995) 329, hep-th/9507012. 
\item [[38]] J. Maldacena , J. Michelson  and  A. Strominger,«  $ADS_3$ Black Holes and Stringy Exclusion Principale »; JHEP 12 (1998) 05 ,hep-th/9908142.
\item [[39]]N. Seiberg and E. Witten, « String Duality and Non Commutative geometry  »; JHEP 9909(1999) 032,  hep-th/9908142.  
\item [[40]]  N. Nekarasov and A. Schwartz, «  Instanton on  Non Commutative $R^4$ and (2,0) Superconformal Six dimensional theory», hep-th/9802068.  
\item [[41]] Shiraz Minwalla, Mark. Van. Raamsdonk and Nathan Seiberg,«   
Non commutative perturbative Dynamics », hep-th/9912072.  
\item [[42]] S. Gukov , C. Vafa and E. Witten ,«  CFT's From Calabi Yau Four-folds »; hep-th/9906070 
\item [[43]] E. Silverstein and E. Witten «  Global U(1) R-symmetry and Conformal Invariance of(0,2) Models», hep-th/9403054  
\item [[44]] P.Aspinwall, B. Green and D.R. Plesser; Nucl Phys B416 414.
\item [[45]] P.Aspinwall and  B.R. Greene; Nucl Phys B437 (1995) 205-230.
\item [[46]] E. Witten, «  Small Instantons In String Theory » ; Nucl . 
Phys B460, 541 (1996),hep-th/9511030.
 \item [[47]] A. Galperin, E. Ivanov, S. Kalitzin, V. Ogievetsky and E. Sokatchev;  Class . Quant . Grav 1(1984) 469. 
\item [[48]] E. Sahraoui and E. H. Saidi,« On the completly integrable four dimensions N=2 hypermultiplet self - couplings»; Class . Quant . Grav 16(1999) 1605. 
\item [[49]] I. Benkadour , M. Bennai, E. Y. Diaf and E. H. Saidi , « On  Matrix  model Compactification on non commutative $F_0$ geometry » ; to appear in  Class . Quant . Grav 17(2000) 
\item [[50]] T. Lhalabi and E.H.Saidi; Inter Jour Mod Phys3 (1988) 187-201 
\item [[51]] E. Sokatchev and K. Stelle; Class . Quant . Grav 4(1987)501.
\item [[52]] k. Schoutens ; Phys .  Rev. lett 79 (1997) 2608 , Cond-mat / 9706166.
\item [[53]] M. Bershadsky , T. M. Chiang , B. Grenne , A. Johansen and  C.I. Lazariou «  F-theory and linear Sigma Models» ; Nucl . Phys . B527(1998) 531-570, hep-th/9712023.  
 \item [[54]]M. Berkooz and H. Verlinde; Matrix Therory, ADS/CFT and Higgs branch, hep-th /9907100   
\item [[55]] A. Belhaj and E.H. Saidi, work in progress.
\item [[56]]  Batyrev,(1994)J.Alg. Geo. 3 493
\end{enumerate}

\end{document}